\documentclass[prc,superscriptaddress,showpacs,psfig,showkeys,nofootinbib]{revtex4}
\usepackage{amsmath}
\usepackage{amssymb}
\usepackage{amsfonts}
\usepackage{graphicx}
\usepackage{epsfig}
\usepackage{dcolumn}
\usepackage{hyperref}
\usepackage{epstopdf}
\usepackage{color}
\usepackage[utf8]{inputenc}
\usepackage{amsthm}
\usepackage{fontenc}
\usepackage{bm}
\RequirePackage{slashed}\usepackage{cancel}
\usepackage[normalem]{ulem}

\usepackage{array,booktabs}
\newcolumntype{C}{>{$\displaystyle}c<{$}}
\usepackage{soul}
\usepackage{color,soul}

\usepackage{ulem}

\begin{document}


\title{Probing fluctuating protons using forward neutrons in soft and hard inelastic proton-nucleus scattering }

\author{M. Alvioli}
\affiliation{Consiglio Nazionale delle Ricerche,
Istituto di Ricerca per la Protezione Idrogeologica,
via Madonna Alta 126, I-06128, Perugia, Italy}
\affiliation{
Istituto Nazionale di Fisica Nucleare, Sezione di Perugia,
via Pascoli 23c, I-06123, Perugia, Italy}

\author{V. Guzey}
\affiliation{University of Jyvaskyla, Department of Physics, P.O. Box 35, FI-40014 University
of Jyvaskyla, Finland and Helsinki Institute of Physics, P.O. Box 64, FI-00014 University of Helsinki, Finland}

\author{M. Strikman}
\affiliation{Pennsylvania  State  University,  University  Park,  PA,  16802,  USA}

\date{\today}

\begin{abstract}

We present a model for the distribution of the number of forward neutrons emitted in soft (minimum bias) and hard inelastic 
proton-nucleus ($pA$) scattering at the LHC.
It is based on the Gribov-Glauber model for the distribution over the number of inelastic collisions
(wounded nucleons) combined with a parametrization of cross section (color) fluctuations in the projectile proton, 
which depend on the parton momentum fraction $x_p$, and the assumption of independent neutron emissions.
We observe that while the effect of soft color fluctuations is modest, a reduction of the interaction strength 
with increasing $x_p$ provides the dominant effect.
It allows us to qualitatively describe the trend of
the ATLAS data on the ZDC energy spectra of forward neutrons emitted in 
dijet production in inelastic $pA$ scattering at $\sqrt{s_{NN}}=8.16$ TeV.

\end{abstract}


\maketitle

\section{Introduction}
\label{sec:intro}

High energy deuteron-nucleus scattering at the Relativistic Heavy Ion Collider (RHIC) and proton-nucleus ($pA$) scattering at the Large Hadron
Collider (LHC)~\cite{Salgado:2011wc,Citron:2018lsq} have played an essential role in studies of quantum chromodynamics (QCD) at small $x$.
Additionally, complementary information is provided by high energy photon-nucleus scattering through ultraperipheral collisions (UPCs) at the RHIC and the LHC, whose full potential is currently actively explored, for reviews, see~\cite{Baltz:2007kq,Contreras:2015dqa,Klein:2019qfb}.
Further, it is anticipated that qualitative improvement in our understanding of small-$x$ dynamics of QCD will be achieved at the planned Electron-Ion Collider (EIC) in USA~\cite{AbdulKhalek:2021gbh}.

Narrowing down the range of outstanding problems, one of the open questions is the mechanism of nuclear shadowing 
(NS) and its possible relation to parton saturation. On the one hand, in hard scattering with nuclei such as deep inelastic scattering (DIS) off fixed nuclear targets and electroweak boson and dijet production in proton-nucleus scattering at the LHC, NS suppresses the nuclear parton
distributions (PDFs) for $x < 0.1$; see~\cite{Klasen:2023uqj} for a recent review.  
This defines the initial conditions (cold nuclear matter effects) for heavy ion scattering. On the other hand, in soft processes involving nuclei, NS probes the space-time picture of high energy scattering and is sensitive to models of the
composite structure of hadronic projectiles. It also illustrates  the important connection between NS and diffraction~\cite{Gribov:1968jf,Frankfurt:2000tya,Frankfurt:2022jns}.

In this article, we focus on a soft mechanism of NS, which is based on the combination of the Gribov-Glauber model for hadron-nucleus scattering with the concept of hadronic (cross section, or color) fluctuations of the proton projectile. Its application to minimum bias 
(MB) inelastic
proton-nucleus scattering at the LHC kinematics allows one to study NS as a function of the number of inelastic proton-nucleon collisions
(wounded nucleons), which encodes information on the transverse geometry of the collision~\cite{Alvioli:2014sba}, \textit{i.e.}, information on the impact parameter dependence of NS. Further, it was suggested in~\cite{Alvioli:2014sba} that imposing the addition condition of the presence of a hard trigger, 
for example, selecting events containing jets with high transverse energy $E_T$
provides a more differential snapshot of the collision and gives an access to
various hadronic fluctuations interacting with different strength. It emphasizes the notion of ``flickering'' of the fluctuating proton. Application of these ideas~\cite{Alvioli:2014eda,Alvioli:2017wou} to inclusive jet production in proton-lead ($p$Pb) scattering at the LHC~\cite{ATLAS:2014cpa} and deuteron-gold ($d$Au) scattering at RHIC~\cite{PHENIX:2015fgy} has given evidence for $x_p$-dependent 
hadronic fluctuations in the proton.
The method consists in considering the configurations containing large-$x_p$ partons, which interact with 
nuclear target nucleons with a cross section that is smaller than the $x_p$-averaged one and, hence, correspond to smaller transverse sizes.
This picture represents an example of the general QCD phenomenon of color transparency (CT)~\cite{Frankfurt:1994hf,Dutta:2012ii}, which manifests itself through 
the ``shrinking of the proton'' at large $x_p$. The latter can be quantified in terms of parton densities in the transverse plane
using the language of generalized parton distributions, see, \textit{e.g.}, \cite{Burkardt:2002hr,Miller:2022kxt,Brodsky:2022bum,Frankfurt:2022tyy}.

We recently extended these ideas to high energy inelastic photon-nucleus ($\gamma A$) scattering in heavy ion UPCs~\cite{Alvioli:2024cmd} (see also~\cite{Alvioli:2016gfo}), which is accompanied 
by emission of forward neutrons from nuclear breakup that are detected by the zero degree calorimeters (ZDC) with high efficiency. Assuming a simple relation between the number of evaporation neutrons produced in the photon-nucleus collision
with the number of wounded nucleons (inelastic collisions), we predicted that the distributions of neutrons detected in a ZDC gives a novel probe
of the mechanism of NS, including its $x$ and impact parameter dependence.
Note that the use of ZDC to learn about the
dynamics of color dipole-nucleon interactions in nuclei using $\gamma A$ scattering was first suggested in~\cite{Strikman:2005ze}. Also,
determination of the electron-nucleus collision geometry using forward neutrons produced in
electron-nucleus DIS at the EIC kinematics and numerical simulations for energy loss, hadron multiplicity, and
dihadron correlations were presented in~\cite{Zheng:2014cha}.

In the present paper, we build on the results of~\cite{Alvioli:2014sba,Alvioli:2014eda,Alvioli:2017wou,Alvioli:2024cmd} and 
suggest a relation between NS, where hadronic fluctuations 
of the proton depend on $x_p$,
with the distribution of forward neutrons in inelastic soft (minimum bias) and hard $p$A scattering at the LHC. It allows us to qualitatively explain the spectrum of energy deposited by forward neutrons in the Pb-going ZDC, $E^{\rm Pb}_{\rm ZDC}$, which was recently
measured in dijet production in inelastic $pA$ scattering at $\sqrt{s_{NN}}=8.16$ TeV by the ATLAS collaboration~\cite{ATLAS:2025hac}.

The paper is organized as follows. In Sec.~\ref{sec:Pnu}, we present the optical limit expressions for the $P(\nu)$ distributions over the number of wounded nucleons 
$\nu$ produced in MB and hard inelastic $pA$ scattering,
including also the effect of cross section (color) fluctuations. The resulting distributions are studied numerically 
using the Monte Carlo Glauber (MCG) approach developed in~\cite{Alver:2008aq,Alvioli:2012pu,Bozek:2019wyr,Lonnblad:2021fyl}
and their comparison is discussed.
One of our main findings is that color fluctuations weakly affect $P(\nu)$.
Influence of the $x_p$ dependence on the proton color fluctuations and the distributions $P(\nu)$ is discussed in Sec.~\ref{sec:cf}.
We observe that it sizably modifies the shape of $P(\nu)$ and
show that with an increase of $x_p$, the peak of $P(\nu)$ shifts toward lower $\nu$.
In Sec.~\ref{sec:fn}, we convert $P(\nu)$ into the $P_{\rm comb}(N)$
distribution of the number of forward neutrons $N$ and discuss its properties.
As an example of application of $P_{\rm comb}(N)$ in the hard trigger case, in Sec.~\ref{sec:E_ZDC}, 
we transform the distribution as a function of $N$
into the normalized ZDC energy spectrum of forward neutrons and compare it with the ATLAS data on dijet production in inelastic $pA$ scattering at $\sqrt{s_{NN}}=8.16$ TeV and show that our model provides a qualitative description of the trends of the data,
with distinctive similar features as well as substantial differences.
Finally, we summarize our results in Sec.~\ref{sec:conclusions}.

\section{Nuclear shadowing, cross section fluctuations of the proton, and the distribution over wounded nucleons}
\label{sec:Pnu}

\subsection{Minimum bias inelastic proton-nucleus scattering}
\label{subsec:MB}

In has been known since late 70s that as a consequence of unitarity of the Glauber theory for high energy hadron-nucleus scattering, when all
the possible inelastic intermediate states between successive scatterings are included, there is a simple relation~\cite{Bertocchi:1976bq} between the average number of
wounded nucleons $\langle \nu \rangle$ and the total inelastic (minimum bias, MB) proton-nucleus cross section $\sigma_{pA}^{MB}$.
Indeed, the cross section $\sigma_{pA}^{MB}$ can be expressed as a sum of the partial cross sections $\sigma_{\nu}$ as follows,
\begin{equation}
\sigma_{pA}^{MB}=\sum_{\nu=1}^{A} \sigma_{\nu} \,,
\label{eq:mb1}
\end{equation}
where in the optical limit approximation, $\sigma_{\nu}$ is given by the following expression,
\begin{equation}
\sigma_{\nu}=\frac{A!}{(A-\nu)!\nu!} \int d^2 {\vec b}\, (\sigma_{\rm in} T_A(\vec{b}))^{\nu}(1-\sigma_{\rm in}T_A(\vec{b}))^{A-\nu} \,.
\label{eq:mb2}
\end{equation}
Here $\sigma_{\rm in}$ is the MB inelastic proton-nucleon cross section; $T_A(b)=\int dz \rho_A(\vec{b},z)$ is the nuclear density in the transverse plane characterized by the impact parameter vector ${\vec b}$ from the center of the nucleus with $\rho_A(\vec{b},z)$ the nuclear density; $A$ is the nucleus mass number. For brevity, we do not explicitly show
the energy dependence of the involved cross sections.

As follows from Eq.~(\ref{eq:mb2}), $\sigma_{\nu}$ corresponds to the physical process, in which 
the projectile proton undergoes inelastic production on $\nu$ nucleons of the target, while the remaining $A-\nu$ nucleons provide inelastic
absorption. Hence, $\nu$ gives the number of inelastic proton-nucleon collisions, which is often referred to as the number of wounded nucleons.
 The average number of proton-nucleon inelastic interactions (wounded nucleons) $\langle \nu \rangle$ is 
\begin{equation}
\langle \nu \rangle =\frac{\sum_{\nu=1}^{A} \nu \sigma_{\nu}} {\sigma_{pA}^{MB}}=
\frac{A \sigma_{\rm in}}{\sigma_{pA}^{MB}} \,.
\label{eq:mb4}
\end{equation}
The last relation is equivalent to Abramovsky-Gribov-Kancheli (AGK) cancellation, which connects the multiplicities of hadron production (inclusive spectra) in inelastic processes with their theoretical classification in terms of the
number of cut reggeon exchanges~\cite{Abramovsky:1973fm}.

It is important to point out that if one quantifies the magnitude of NS in inelastic proton-nucleus scattering by the ratio $R_{pA}=\sigma_{pA}^{MB}/(A \sigma_{\rm in})$, Eq.~(\ref{eq:mb4}) states that $R_{pA}$ is inversely proportional to $\langle \nu \rangle$,
\begin{equation}
R_{pA}=\frac{1}{\langle \nu \rangle}  \,,
\label{eq:mb4b}
\end{equation}
which in principle gives an additional constraint on NS.

It is useful to introduce the distribution of the number of wounded nucleons $P(\nu)$, 
\begin{equation}
P(\nu)=\frac{\sigma_{\nu}}{\sum_{\nu=1}^{A} \sigma_{\nu}}= \frac{\sigma_{\nu}}{\sigma_{pA}^{MB}} \,.
\label{eq:mb3}
\end{equation}
Note that $\sum_{\nu} P(\nu)=1$ and $\sum_{\nu} \nu P(\nu)=\langle \nu \rangle$.
Since the distribution $P(\nu)$ is a more differential quantity than $\langle \nu \rangle$, it contains more detailed information on NS compared to the $\nu$-averaged case, see Eq.~(\ref{eq:mb4b}). It is one of the goals of this work to substantiate this claim.

Equations~(\ref{eq:mb1}) and (\ref{eq:mb2}) assume that the projectile proton interacts with a nuclear target as a single state with a well-defined cross section. The space-time picture of the strong interaction, which is more appropriate at high energies, is to view 
the incoming proton as a superposition of eigenstates of the scattering operator corresponding in general to different
cross sections~\cite{Good:1960ba}. The resulting method, which is often referred to in the literature as the Good-Walker formalism,
provides an economical description of diffraction and, in particular, explains the origin of diffractive dissociation 
of beam particles. It is important to note that these diffractive final states 
correspond to inelastic intermediate states, which build up the shadowing correction in the Gribov-Glauber model of NS.

In this formalism, the values of the projectile cross section fluctuate across individual events.
Since the interaction strength is determined by the underlying QCD color interactions, we use the terms
cross section and color fluctuations interchangeably. 
The effect of (color, cross section) fluctuations (CF) of the projectile proton can be modeled by introducing the distribution $P_p(\sigma)$, where 
$\sigma$ is the cross section for a given hadronic fluctuation~\cite{Frankfurt:2022jns,Miettinen:1978jb,Blaettel:1993ah}. 
Since we require the inelastic rather than the total cross section in Eq.~(\ref{eq:mb2}), one needs to rescale $P_p(\sigma)$, which is 
designed for the total cross section, by the factor of $\sigma_{\rm in}/\sigma_{\rm tot}$, where $\sigma_{\rm tot}$ 
and $\sigma_{\rm in}$
are the total and inelastic photon-nucleon
cross sections. 
Thus, the inclusion of CF effects
results in the following expression for the partial cross section, cf.~Eq.~(\ref{eq:mb2}),
\begin{equation}
\sigma_{\nu}^{\rm CF}=\frac{A!}{(A-\nu)!\nu!}\int d\sigma \frac{\sigma_{\rm in}}{\sigma_{\rm tot}} P_p(\sigma)  \int d^2 {\vec b} \, (\sigma T_A(\vec{b}))^{\nu}(1-\sigma T_A(\vec{b}))^{A-\nu} \,.
\label{eq:mb2_cf}
\end{equation}
The corresponding distribution of the number of wounded nucleons can then be defined as follows [compare to Eq.~(\ref{eq:mb3})],
\begin{equation}
P_{\rm CF}(\nu)=\frac{\sigma_{\nu}^{\rm CF}}{\sum_{\nu=1}^{A} \sigma_{\nu}^{\rm CF}} \,.
\label{eq:P_CF}
\end{equation}
Since the distribution $P_p(\sigma)$ is peaked around $\sigma_{\rm tot}$, CF do not significantly affect the shape of the distribution $P_{\rm CF}(\nu)$;
the most noticeable effect is a significant enhancement of $P_{\rm CF}(\nu)$ for large $\nu > 20$, where 
both $P(\nu)$ and $P_{\rm CF}(\nu)$ are
 very small~\cite{Alvioli:2014sba}.
 Note that on a logarithmic scale, $P_{\rm CF}(\nu)$ is an order of magnitudes larger than $P(\nu)$, which can make a difference
in some cases; see Fig.~3 of \cite{Alvioli:2013vk}.

\subsection{Inelastic proton-nucleus scattering in presence of a hard trigger}
\label{subsec:HT}

Triggering on a particular hard process with a small cross section $\hat{\sigma}_0 \sim 1/p_T^2 \ll \sigma_{\rm in}$, where $p_T$ is 
a characteristic large scale, is equivalent to selecting the inelastic events, which lead to a given final state produced in this hard process.
It corresponds to the physical situation, in which one expands the MB partial cross section $\sigma_{\nu}$ in powers of $\hat{\sigma}_0$ keeping only
the first term.
Operationally, it amounts to the following substitution of the term $(\sigma_{\rm in} T(b))^{\nu}$ in Eq.~(\ref{eq:mb2}),
\begin{equation}
(\sigma_{\rm in} T_A(\vec{b}))^{\nu} \to \nu \hat{\sigma}_0 T_A(\vec{b}) (\sigma_{\rm in} T_A(\vec{b}))^{\nu-1}=\frac{\nu \hat{\sigma}_0}{\sigma_{\rm in}}  (\sigma_{\rm in} T(b))^{\nu} \,.
\label{eq:ht1}
\end{equation}
One can then define the partial cross section in the presence of a hard trigger (HT) $\sigma_{\nu}^{\rm HT}$,
\begin{eqnarray}
\sigma_{\nu}^{\rm HT} &=& \frac{A!}{(A-\nu)!\nu!} \frac{\nu \hat{\sigma}_0}{\sigma_{\rm in}} \int d^2 {\vec b} \, (\sigma_{\rm in} T_A(\vec{b}))^{\nu}(1-\sigma_{\rm in}T_A(\vec{b}))^{A-\nu} \nonumber\\
&=& \frac{\nu \hat{\sigma}_0}{\sigma_{\rm in}} \sigma_{\nu} \,.
\label{eq:ht2}
\end{eqnarray}
Note that in a similar form, a relation between $\sigma_{\nu}^{\rm HT}$ and $\sigma_{\nu}$ was suggested in~\cite{Alvioli:2017wou}.

The corresponding distribution of the number of wounded nucleons is
\begin{equation}
P_{\rm HT}(\nu)=\frac{\sigma_{\nu}^{\rm HT}}{\sum_{\nu=1}^{A} \sigma_{\nu}^{\rm HT}}=\frac{\nu \sigma_{\nu}}{\sum_{\nu=1}^{A} \nu \sigma_{\nu}}=\frac{\nu \sigma_{\nu}}{A \sigma_{\rm in}} \,.
\label{eq:ht3}
\end{equation}
A comparison with Eq.~(\ref{eq:mb3}) shows that the presence of a hard process dramatically affects the shape of $P_{\rm HT}(\nu)$:
an extra power of $\nu$ compared to $P(\nu)$ suppresses the contribution of small $\nu$ and enhances the contribution of intermediate and large $\nu$.

Finally, as in the case of the MB inelastic proton-nucleus scattering considered above, one can generalize the partial
cross section in the presence of a hard trigger to include the effect of cross section fluctuations. The corresponding 
partial cross section reads 
\begin{equation}
\sigma_{\nu}^{\rm HT+CF} = \frac{A!}{(A-\nu)!\nu!} \frac{\nu \hat{\sigma}_0}{\sigma_{\rm in}} \int d\sigma \frac{\sigma_{\rm in}}{\sigma_{\rm tot}} P_p(\sigma) \int d^2 {\vec b} \,
 (\sigma T_A(\vec{b}))^{\nu}(1-\sigma T_A(\vec{b}))^{A-\nu}  \,.
\label{eq:ht2_cf}
\end{equation}
One can see that the expression for $\sigma_{\nu}^{\rm HT+CF}$ is obtained by combining Eqs.~(\ref{eq:mb2_cf}) and (\ref{eq:ht2}).
The distribution of the number of wounded nucleons (inelastic collisions) corresponding to $\sigma_{\nu}^{\rm HT+CF}$ is
\begin{equation}
P_{\rm HT+CF}(\nu)=\frac{\sigma_{\nu}^{\rm HT+CF}}{\sum_{\nu=1}^{A} \sigma_{\nu}^{\rm HT+CF}} \,.
\label{eq:ht2_cf2}
\end{equation}

\subsection{Numerical results for the distributions over the number of wounded nucleons}
\label{subsec:numerics}

It is important to emphasize that the expressions in Eqs.~(\ref{eq:mb1})--(\ref{eq:ht2_cf2}) represent the 
optical limit approximation to the quantities investigated here, which serve as a useful illustration of the $pA$ process in terms 
of the nuclear density $\rho_A(\vec{b},z)$ and the nuclear thickness function $T_A(\vec{b}) = \int dz \rho_A(\vec{b},z)$.
The actual numerical results presented in all the figures in this work are based on the calculation 
performed within the Monte Carlo Glauber (MCG) approach~\cite{Alver:2008aq,Alvioli:2012pu,Bozek:2019wyr,Lonnblad:2021fyl}.
The MCG model replaces the smoothed-density description
of the target nucleus by that based on individual nucleons placed at discrete positions in the three-dimensional space,
with nuclear configurations prepared with a Monte Carlo realization. Configurations can be prepared including
nucleon-nucleon (NN) correlations~\cite{Alvioli:2014sba,Alvioli:2009ab}, nuclear deformations~\cite{Hammelmann:2019vwd}, and neutron skin effects~\cite{Alvioli:2018jls}. 

The advantage is that MCG permits an event-by-event simulation, in which each projectile-target
pair experiences inelastic (soft) collisions based on the value of the total proton-nucleon cross section $\sigma_{\rm tot}$
and the proton-nucleon distance in the transverse plane $\vec{s}$.
The probability of the interaction is given as a function of $\vec{s}$
by $1 - [1 - \Gamma(\vec{s})]$, 
with $\Gamma(\vec{s})$ defined as follows 
\begin{equation}
\Gamma(\vec{s}) = \frac{\sigma_{\rm tot}}{4\pi B} \exp(-|\vec{s}|^2/2B) \,,
\label{eq:dop1}
\end{equation}
where $B$ is the slope of the $t$ dependence of the elastic proton-nucleon cross section. In this work,
we use $\sigma_{\rm tot}=102$ mb and $B=\sigma_{\rm tot}^2/(16 \pi \sigma_{\rm el})=20$ GeV$^{-2}$ corresponding to  
$\sqrt{s_{NN}}=8.16$ TeV~\cite{ParticleDataGroup:2018ovx}.

In each simulated event, the MCG approach allows one to distinguish wounded nucleons (participants)
in the target nucleus and spectators nucleons~\cite{Alvioli:2010yk}. The MCG model also permits
a simple implementation of color fluctuations, by selecting in each simulated event a specific
value of the $NN$ cross section probed randomly according to the distribution of Eq.~(\ref{eq:Psigma}) below; this
represents a particular ``frozen'' configuration of the projectile. Moreover, in the MCG descriptions, one
has a full impact parameter dependence of the process, which allows for a simple definition of
centrality classes, ultimately related to particle multiplicity in experimental analyses.

In addition, the MCG approach allowed one to use the hard trigger mechanism for $pA$ collisions~\cite{Alvioli:2014sba} and subsequently 
extend it to $dA$~\cite{Alvioli:2014eda} and double
parton interactions~\cite{Alvioli:2019kcy}. In this approach, for each global impact parameter,
the hard interaction occurs at any point in the transverse plane with the probability $P_h(\vec{\rho})$
given by superposition of the gluon transverse distribution in target nucleons, 
\begin{equation}
P_h(\vec{\rho})=\frac{1}{\pi B_h^2}\exp(-|\vec{\rho}|^2/B_h^2) \,,
\label{eq:P_h}
\end{equation}
 where $\vec{\rho}$ is the transverse separation
between the global impact parameter and the hard interaction point and $B_h$ = 0.5 fm. In each
event, we integrate over the position of the hard interaction point in the transverse plane with
the probability of each interaction point $P_h(\vec{\rho})$ as a weight.

The distribution $P_p(\sigma)$ introduced in Sec.~\ref{subsec:MB} cannot be calculated from the first principles of QCD. 
However, its shape can be constrained by its first few moments as follows.
Assuming that $P_p(\sigma)$ can be parametrized as~\cite{Frankfurt:2022jns,Blaettel:1993ah}, 
\begin{equation}
P_p(\sigma)=N_p \frac{\sigma}{\sigma+\sigma_0} e^{-(\sigma-\sigma_0)^2/(\Omega \sigma_0)^2} \,,
\label{eq:Psigma}
\end{equation}
the three free parameters $N_p$, $\sigma_0$, and $\Omega$ are determined using the following constraints,
\begin{eqnarray}
\int d\sigma P_p(\sigma) & =& 1 \,, \nonumber\\
\int d\sigma P_p(\sigma) \sigma & =& \sigma_{\rm tot} \,, \nonumber\\
\int d\sigma P_p(\sigma) \sigma^2 & =& \sigma_{\rm tot}^2 (1+\omega_{\sigma}) \,.
\label{eq:Psigma2}
\end{eqnarray}
Here $\sigma_{\rm tot}$ is the total proton-nucleon cross section, and $\omega_{\sigma}$ quantifies the dispersion of $P_p(\sigma)$ around its peak, textit{i.e.}, it controls its width. 
Both $\sigma_{\rm tot}$ and $\omega_{\sigma}$, as well as $N_p$, $\sigma_0$, and $\Omega$, depend on the collision energy. The results presented below correspond to the center of mass energy 
$\sqrt{s_{NN}}=8.16$ TeV, where $\sigma_{\rm tot}=102$ mb and $\sigma_{\rm in}=75$ mb~\cite{ParticleDataGroup:2018ovx},  and 
$\omega_{\sigma}=0.1$~\cite{Guzey:2005tk}.

\begin{figure}[t!]
  \centerline{%
    \includegraphics[width=9cm]{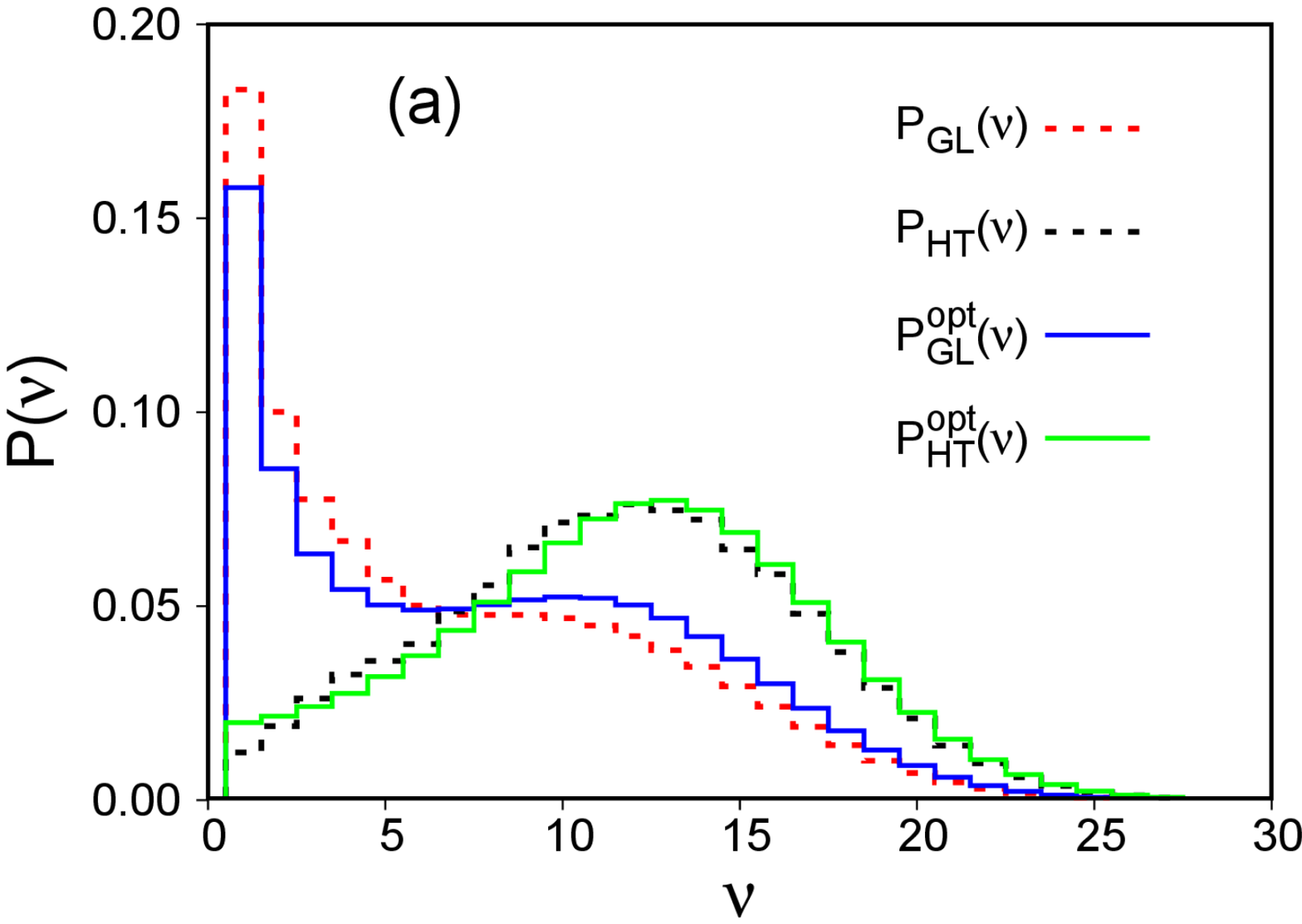}
    \includegraphics[width=9cm]{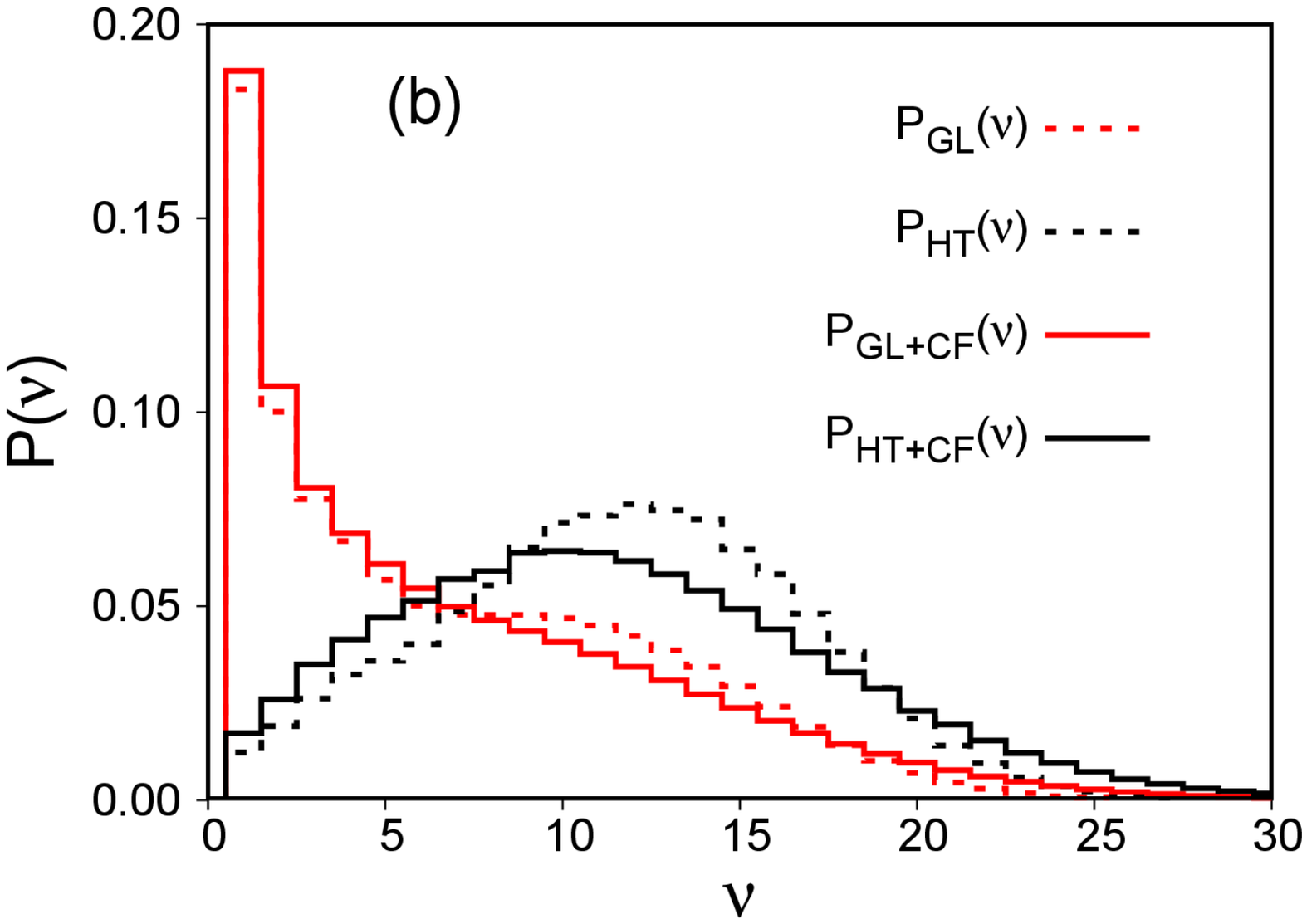}}
  \caption{The distributions $P(\nu)$ of the number of wounded nucleons $\nu$ at $\sqrt{s_{NN}}=8.16$ TeV.
  (a) Comparison of the optical limit $P^{\rm opt}_{\rm GL,HT}(\nu)$ and MCG $P_{\rm GL,HT}(\nu)$ results
  in the MB and HT cases.
   (b) The MCG results for the minimum bias $P_{\rm GL}(\nu)$ (red dashed line), the minimum bias with CF $P_{\rm GL+CF}(\nu)$ 
   (red solid line), the hard trigger $P_{\rm HT}(\nu)$ (black dashed line), and the hard trigger with CF $P_{\rm HT+CF}(\nu)$ 
  (black solid line).}
  \label{fig:P_nu_fig1}
\end{figure}

Figure~\ref{fig:P_nu_fig1} presents our predictions for the distribution $P(\nu)$ of the number of wounded nucleons $\nu$ at $\sqrt{s_{NN}}=8.16$ TeV.
Panel (a) compares the results of the optical limit and MCG calculations, which are given by $P^{\rm opt}_{\rm GL,HT}(\nu)$
and $P_{\rm GL,HT}(\nu)$, respectively. The blue solid line corresponds to the MB case and Eq.~(\ref{eq:mb3}),
the green solid line corresponds to the hard trigger case and Eq.~(\ref{eq:ht3}); their MCG counterparts are given by the red and
black dashed lines, respectively. 
One can see from this panel that the optical limit and MCG results are very close, especially
in the HT case. It demonstrates that the optical limit expressions presented in Sec.~\ref{subsec:MB} and \ref{subsec:HT} correctly capture
the nuclear geometry involved in the description of soft and hard $pA$ scattering.

Note that throughout this work, we use the terms ``minimum bias'' and ``Glauber'' interchangeably because they correspond to
the same physical situation, which is illustrated by Eqs.~(\ref{eq:mb1})--(\ref{eq:P_CF}) in Sec.~\ref{subsec:MB}.

Panel (b) of Fig.~\ref{fig:P_nu_fig1} presents $P(\nu)$ calculated using the MCG framework in the four cases considered above: the minimum bias $P_{\rm GL}(\nu)$ as the red dashed line, the minimum bias with cross section fluctuations 
$P_{\rm GL+CF}(\nu)$ as the red solid line, the hard trigger $P_{\rm HT}(\nu)$ as the black dashed line, and the hard trigger with CF 
$P_{\rm HT+CF}(\nu)$ as the black solid line.
One can see from the figure that CF do not dramatically alter the shape of $P(\nu)$ and only enhance the large-$\nu$ tail, where 
$P(\nu)$ is very small. At the same time, the condition to have a hard trigger dramatically changes $P(\nu)$ by suppressing the contribution of small $\nu < 5$ and enhancing the contributions of intermediate and large $\nu > 10$.

\section{Color fluctuations depending on $x_p$}
\label{sec:cf}

The cross section (color) fluctuations introduced in Sec.~\ref{sec:Pnu} account for the overall composite structure of the projectile 
proton, which can be probed in proton-nucleus scattering.
Its description uses the concepts of soft QCD and does not involve the language of partons. On the other hand, hard processes probe the  
dynamics of QCD and, as a result, CF should also reflect the underlying partonic structure of the proton. 
One of its manifestations is the phenomenon of color transparency supported by
the observation that small-size partonic configurations
interact with hadronic targets with reduced cross sections~\cite{Frankfurt:1994hf,Dutta:2012ii}.

This leads to the notion that the proton ``shrinks'' as $x_p$ is increased,
which can be formalized using parton densities in the transverse plane
and the concept of generalized parton distributions~\cite{Burkardt:2002hr,Miller:2022kxt,Brodsky:2022bum,Frankfurt:2022tyy}.
This idea was realized 
in~\cite{Alvioli:2014eda,Alvioli:2017wou}, where it was shown that the data on production of jets with high transverse energy $E_T$ in
inelastic deuteron-nucleus scattering at RHIC and proton-nucleus scattering at the LHC gives an access to the dependence of 
CF on the longitudinal momentum fraction $x_p$ of the active parton in the proton. The emerging physical picture is that partonic configurations in the proton, which are associated with larger $x_p$, interact with a nuclear target with a smaller-than-average
 cross section. 

In this section, we introduce the $x_p$ dependence in our model of CF and examine its effect on 
the distribution of the number of wounded nucleons.
Following~\cite{Alvioli:2014eda,Alvioli:2017wou}, we characterize the $x_p$ dependence of the interaction strength by the parameter
$\lambda(x_p)$,
\begin{equation}
\lambda(x_p)=\frac{\langle \sigma(x_p)\rangle}{\sigma_{\rm tot}} \,,
\label{eq:lambda}
\end{equation}
where 
\begin{equation}
\langle \sigma(x_p)\rangle=\int d\sigma P_p(\sigma,x_p) \sigma \,.
\label{eq:lambda2}
\end{equation}
Operationally, Eqs.~(\ref{eq:lambda}) and (\ref{eq:lambda2}) mean that for a given $\lambda(x_p)$, one finds $\langle \sigma(x_p)\rangle$ using Eq.~(\ref{eq:lambda}).
Substituting it in Eq.~(\ref{eq:lambda2}), one iteratively finds the $x_p$-dependent distribution $P_p(\sigma,x_p)$,
whose shape is given by Eq.~(\ref{eq:Psigma}); it results in CF depending on $x_p$. Note that we keep $\omega_{\sigma}=0.1$ 
in the whole $x_p$-range so that the system of 
equations~(\ref{eq:Psigma2}) can be solved for $N_p$, $\sigma_0$, and $\Omega$
without additional assumptions.

The analyses of~\cite{Alvioli:2014eda,Alvioli:2017wou} suggested that $\lambda(x_p)$ decreases with increasing $x_p$
 in the studied range of intermediate-to-large values of $x_p$, $0.1 < x_p < 0.7$. This region only partially overlaps with the
 $x_p$ range covered by the ATLAS measurement~\cite{ATLAS:2025hac}. Extrapolating the result of Ref.~\cite{Alvioli:2017wou} to small $x_p$, 
 we propose the following reference values for the three $x_p$-bins of the ATLAS data,
\begin{eqnarray}
\lambda(3.3 \times 10^{-1} < x_p < 4.8 \times 10^{-1}) &=& 0.8 \,, \nonumber\\
\lambda(3.6 \times 10^{-2} < x_p < 5.2 \times 10^{-2}) &=& 0.9 \,, \nonumber\\
\lambda(2.8 \times 10^{-3} < x_p < 4.0 \times 10^{-3}) &=& 1 \,.
\label{eq:lambda3}
\end{eqnarray}

\begin{figure}[t!]
  \centerline{%
\includegraphics[width=9cm]{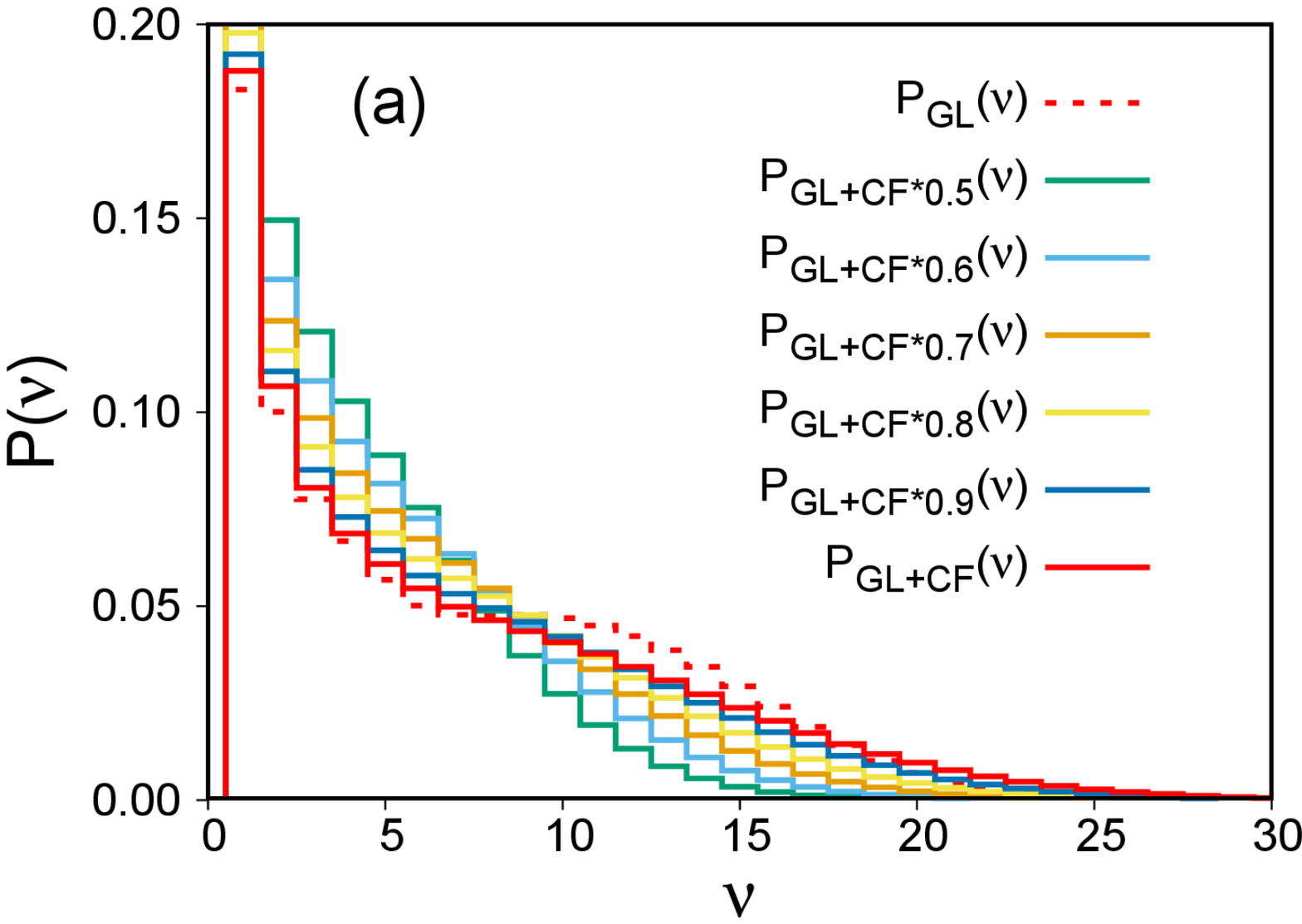}
\includegraphics[width=9cm]{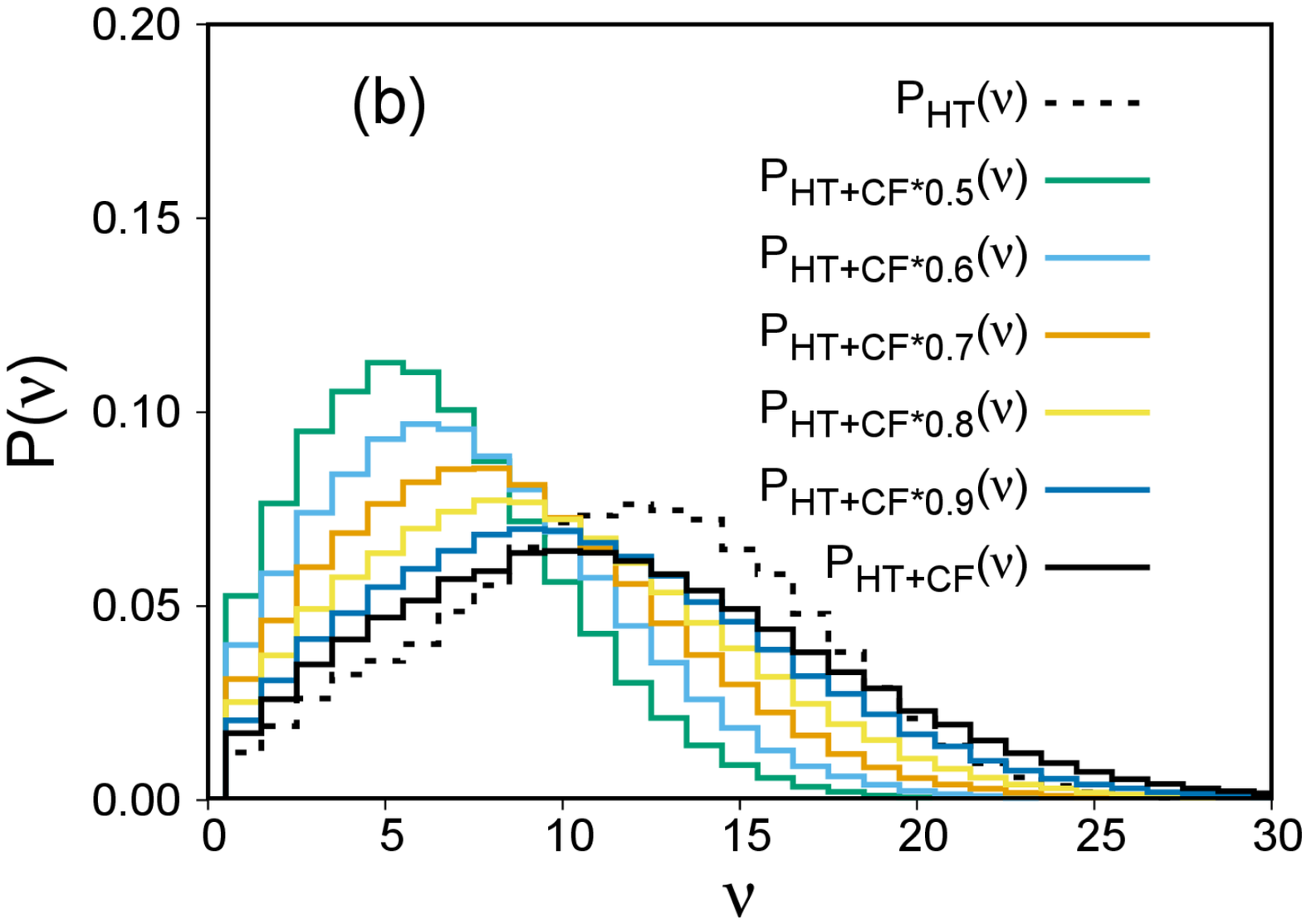}}
  \caption{The distribution $P(\nu)$ of the number of wounded nucleons $\nu$ at $\sqrt{s_{NN}}=8.16$ TeV for different scenarios of the CF dependence on $x_p$ through the parameter $\lambda(x_p)$ in the range $0.5 \leq \lambda(x_p) \leq 1$. 
 Panels (a) and (b) show $P_{\rm CF}(\nu)$ and $P_{\rm HT+CF}(\nu)$, respectively. The dashed curves show the respective 
 results without CF.}
  \label{fig:Pnu_fig23}
\end{figure}

Figure~\ref{fig:Pnu_fig23} presents the distributions of the number of wounded nucleons $\nu$ at $\sqrt{s_{NN}}=8.16$ TeV, where
CF depend on $x_p$. 
Panel (a) shows 
the minimum bias with CF distribution $P_{\rm CF}(\nu)$, while panel (b) gives the 
hard trigger with CF distribution $P_{\rm HT+CF}(\nu)$. Different curves correspond to different values of $\lambda(x_p)$ in the interval
$0.5 \leq \lambda(x_p) \leq 1$; they were
 obtained using the distribution $P_p(\sigma,x_p)$, whose determination procedure is outlined above.
For a comparison, the figure also shows the results without the CF effect by the dashed curves, 
for the calculation of both the minimum bias and hard trigger results.

One can see from Fig.~\ref{fig:Pnu_fig23}
that a decrease of $\lambda(x_p)$ shifts the strength of $P(\nu)$ toward lower $\nu$ and depletes the region of
large $\nu$. This trend can be readily understood by noticing that smaller $\langle \sigma(x_p)\rangle$ correspond to a weaker NS and, hence, to a smaller average number of wounded nucleons, see Eq.~(\ref{eq:mb4b}).
The effect is more pronounced for $P_{\rm HT+CF}(\nu)$ than for $P_{\rm CF}(\nu)$ because the former is more sensitive to hard scattering
kinematics due to its enhanced dependence on the collision geometry via $T_A(\vec{b})$; see Eq.~(\ref{eq:ht1}).
It is actually one of motivations of the ATLAS analysis~\cite{ATLAS:2025hac} to study the sensitivity of event geometry estimators
to the initial state kinematics of the hard scattering in $p$Pb collisions.

\section{Distribution of the number of forward neutrons in ZDC}
\label{sec:fn}

Under certain simplifying assumptions~\cite{Alvioli:2024cmd}, the number of wounded nucleons $\nu$ can be related to the number of forward neutrons $N$ produced
in inelastic scattering with target nucleons. 
Since these neutrons carry the full beam momentum, this picture translates into the
energy deposition in zero degree calorimeters (ZDC) used to detect the neutrons.

In the following, we use the model introduced in Ref.~\cite{Alvioli:2024cmd}, which assumes that each 
proton-nucleon scattering results in the
creation of $\langle M_n \rangle$ neutrons on average, independently of other interactions.
Therefore, the distribution of the number of forward neutrons $N$ can be written as the following convolution,
\begin{equation}
P_{\rm comb}(N) = \sum_{\nu=1}^A P(\nu) P_{\rm Poisson}(N; \nu \langle M_n \rangle) \,,
\label{eq:fn1}
\end{equation}
where $P_{\rm Poisson}(N; \nu \langle M_n \rangle)$ is the Poisson distribution,
\begin{equation}
P_{\rm Poisson}(N; \nu \langle M_n \rangle)=\frac{(\nu \langle M_n \rangle)^N e^{\nu \langle M_n \rangle}}{N!} \,.
\label{eq:fn2}
\end{equation}
The average neutron multiplicity $\langle M_n \rangle$ can be estimated using the analysis of muon-nucleus DIS in coincidence with detection of slow neutrons, $\mu^{-} + A \to n + X$, which has showed that $\langle M_n \rangle \approx 5$ for a lead target~\cite{E665:1995utr}.
It is important to point out that since these data correspond to the average nucleus momentum fraction 
$\langle x_A \rangle =0.015$, where the effect of nuclear shadowing is small, they correspond to $\langle \nu \rangle \approx 1$.
It allows us to correlate each inelastic interaction (wounded nucleon) with a release of 5 forward neutrons.
Note that a similar estimate has been found in~\cite{Zheng:2014cha}.

\begin{figure}[t!]
  \centerline{%
\includegraphics[width=9cm]{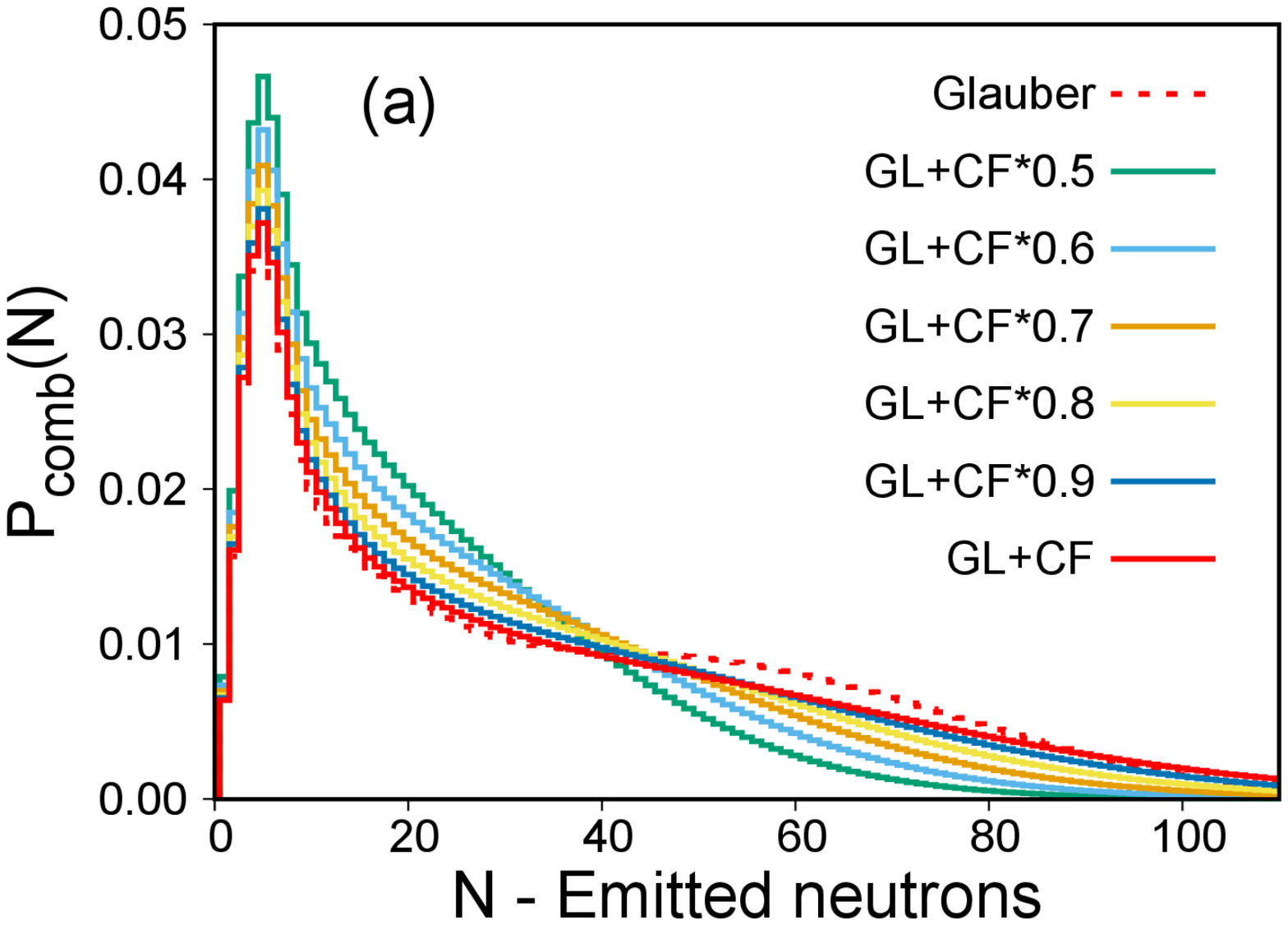}
\includegraphics[width=9cm]{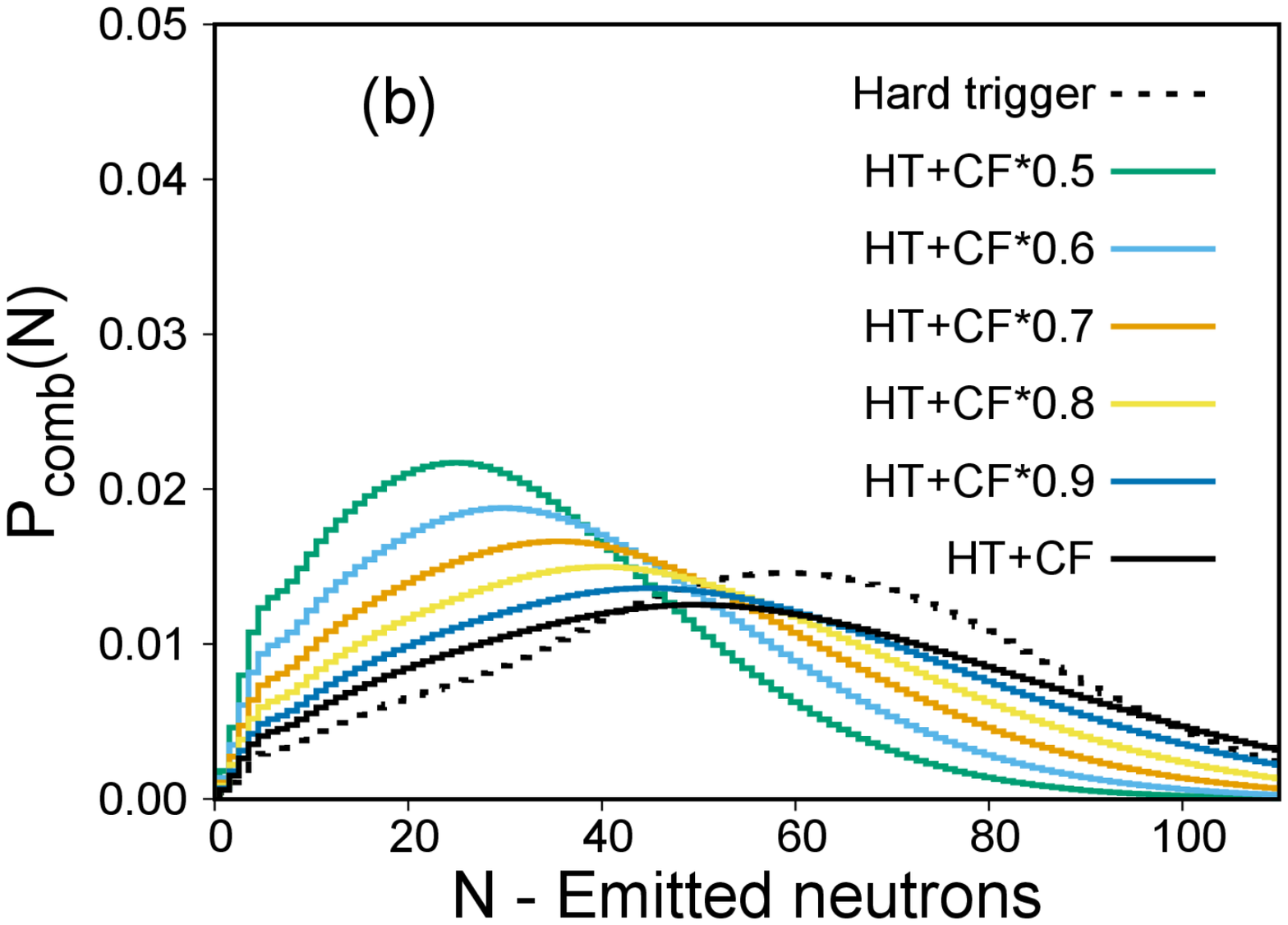}}
  \caption{The distribution $P_{\rm comb}(N)$ of the number of forward neutrons $N$ emitted in inelastic $pA$ scattering
  at $\sqrt{s_{NN}}=8.16$ TeV. Different curves correspond to various scenarios for $P(\nu)$, see Fig.~\ref{fig:Pnu_fig23}, 
  combined with the Poisson distribution for evaporation neutrons, see Eqs.~(\ref{eq:fn1}) and (\ref{eq:fn2}).
 Panels (a) and (b) correspond to the minimum bias and hard trigger cases, respectively. The dashed curves show the respective 
 results without CF.}
  \label{fig:P_comb}
\end{figure}

The resulting distribution $P_{\rm comb}(N)$ as a function of the number of emitted forward neutrons $N$ is shown in Fig.~\ref{fig:P_comb}.
Panel (a) corresponds the minimum bias inelastic $pA$ scattering, while panel (b) includes the condition of a hard trigger.
Different curves correspond to various scenarios for $P(\nu)$, see Fig.~\ref{fig:Pnu_fig23}, combined with the Poisson distribution, see
Eqs.~(\ref{eq:fn1}) and (\ref{eq:fn2}); this is reflected in labeling of the curves. One can see from the figure that the shapes of 
$P_{\rm comb}(N)$ inherit those of $P(\nu)$: CF insignificantly modify them compared to the Glauber model result in the shown interval of $N$. By contrast,
the dependence of CF on $x_p$ noticeably affects
 them by shifting the maxima of the distributions toward lower $N$; the effect 
is especially pronounced in the case of a hard trigger, see panel (b) of Fig.~\ref{fig:P_comb}.

\section{Forward neutron energy distribution in ZDC and comparison to ATLAS data}
\label{sec:E_ZDC}

\begin{figure}[t!]
  \centerline{%
 \includegraphics[width=9cm]{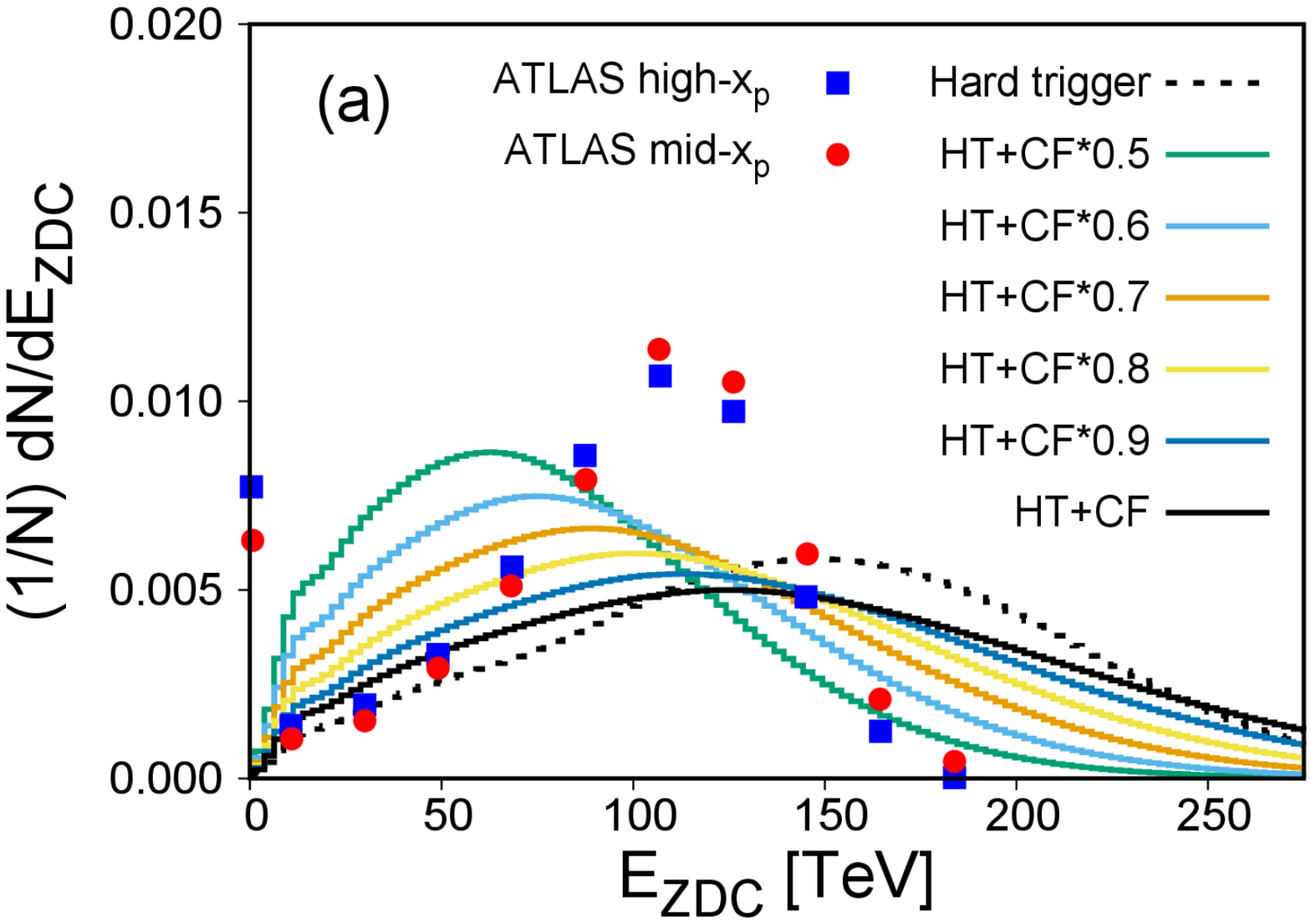}
 \includegraphics[width=9cm]{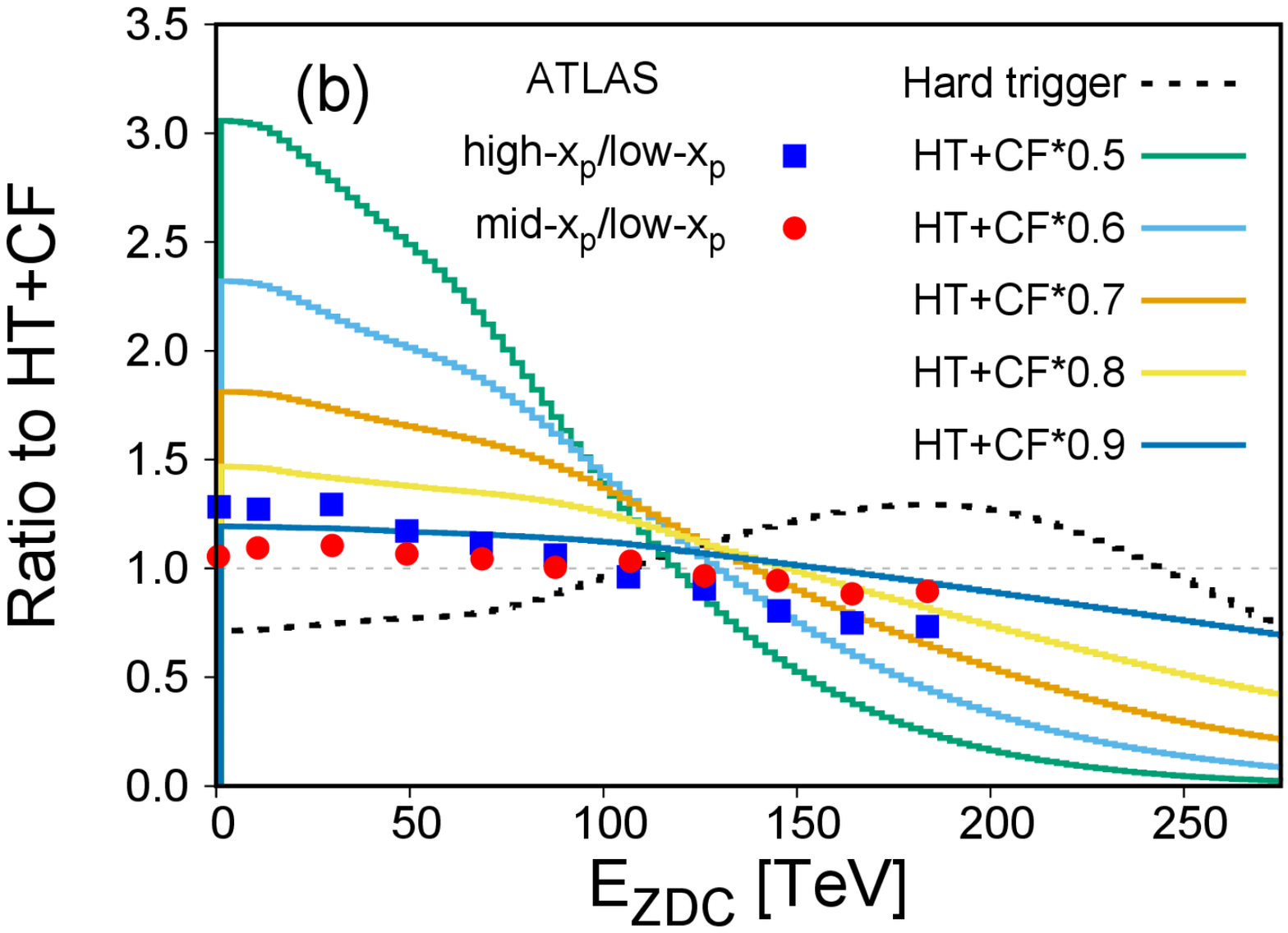}}
  \caption{(a) The normalized ZDC energy spectrum of forward neutrons produced in inelastic proton-lead scattering with dijets at $\sqrt{s_{NN}}=8.16$ TeV, see Eq.~(\ref{eq:fn3}). Different scenarios for the CF dependence on $x_p$ are compared with the 
  preliminary ATLAS data~\cite{ATLAS:2025hac} for mid-$x_p$ (red circles) and high-$x_p$ (blue squares).
  (b) The ratio of the ZDC energy spectra with respect to the ``HT+CF'' case. The ATLAS data~\cite{ATLAS:2025hac} for the ratio between the mid-$x_p$ and high-$x_p$ selection
over the low-$x_p$ selection are shown by the red circles and blue squares, respectively. }
  \label{fig:P_ZDC}
\end{figure}

The distribution of the number of forward neutrons presented in Sec.~\ref{sec:fn} can be converted into the distribution of the neutron energy deposited in the Pb-going zero degree calorimeter (ZDC) $E_{\rm ZDC}$. Assuming that each neutron carries $E_N=2.51$ TeV of energy, one obtains for the normalized distribution of the neutron energy,
\begin{equation}
(1/N) dN/dE_{\rm ZDC} = \frac{P_{\rm comb}(E_N \times N)}{E_N} \,.
\label{eq:fn3}
\end{equation}
This distribution is shown in panel (a) of Fig.~\ref{fig:P_ZDC}, where different curves correspond to the various cases presented in Fig.~\ref{fig:P_comb}, panel (b).
The figure compares the theoretical predictions for $P_{\rm comb}(E_N \times N)/E_N$
with the preliminary ATLAS data on the normalized energy spectrum $(1/N) dN/dE_{\rm ZDC}$ of forward neutrons produced in inelastic proton-lead scattering with dijets at $\sqrt{s_{NN}}=8.16$ TeV~\cite{ATLAS:2025hac}, which are shown by the red circles and the blue squares.
The measurement of the dijet final state kinematics allowed one to reconstruct the momentum fraction $x_p$ of the active parton
in the projective proton and, as a result, to present the data in three bins of $x_p$: $2.8 \times 10^{-3} < x_p < 4.0 \times 10^{-3}$ (low $x_p$), 
$3.6 \times 10^{-2} < x_p < 5.2 \times 10^{-2}$ (mid $x_p$), and $0.33 < x_p < 0.48$ (high $x_p$).
The red circles correspond to mid-$x_p$, and the blue squares are the high-$x_p$ ATLAS data.
Note that the low-$x_p$ and mid-$x_p$ spectra largely overlap within experimental errors and, hence, we do not show 
separately the low-$x_p$ data points. 
Moreover, we do not show the experimental uncertainty bands and errors because our aim is to provide a
qualitative description of the data rather than an accurate explanation.

One can see from Fig.~\ref{fig:P_ZDC} that our model captures the trend of 
the measured energy spectrum: it has a bell-shaped peak
around $E_{\rm ZDC}=100$ TeV. The peak position can be qualitatively understood as follows.  At the considered energy, the most probable number of inelastic proton-nucleon interactions (wounded nucleons) in the presence of a hard trigger is $\nu \approx 10$, see Fig.~\ref{fig:P_nu_fig1}. Since each of such interactions releases on average $\langle M_n \rangle=5$ neutrons, each of which carrying $E_N=2.51$ TeV, one readily obtains the peak position quoted above.
  
We explained 
in Sec.~\ref{sec:cf} that the parameter $\lambda(x_p)$ decreases as $x_p$ increases. In the kinematics of the ATLAS measurement, the mid-$x_p$ bin corresponds to $\lambda(x_p) \approx 0.9$ (the blue curve), and the high-$x_p$ region corresponds to $\lambda(x_p) =0.8$ (the yellow curve). While our model provides an adequate description of the ATLAS data for 
$E_{\rm ZDC} < 100$ TeV, it underestimates the height of the peak around $E_{\rm ZDC}=100$ TeV by approximately a factor of two, and it 
does not reproduce the strong suppression of the spectrum for $E_{\rm ZDC} > 150$ TeV.
A common feature of our theoretical estimate and the data is that in the studied range of $x_p$, the sensitivity to the CF dependence on $x_p$ 
is weak.

One can also consider the ratio of the forward neutron energy spectra at different $x_p$. 
Panel (b) of Fig.~\ref{fig:P_ZDC} shows the ratios of the energy spectra shown in panel (a) of this figure to the theoretical prediction
in the ``HT+CF'' case (the back curve). They are compared with the ATLAS data for the ratio between the mid-$x_p$ and high-$x_p$ selection
over the low-$x_p$ selection~\cite{ATLAS:2025hac}, which are labeled by the red circles and blue 
squares, respectively.
One can see from this panel that our model captures the general trend of the data, where the ratio in question is enhanced above unity for 
$E_{\rm ZDC} < 100$ TeV and dips below unity for $E_{\rm ZDC} > 100$ TeV; the effect increases with the increase of $x_p$.
As discussed in Sec.~\ref{sec:cf}, the range of $x_p$ covered by the ATLAS data corresponds to $0.8 \leq \lambda(x_p) \leq 1$, where the dependence of color fluctuations on $x_p$ is expected to be weak. This is illustrated in panel (b) of Fig.~\ref{fig:P_ZDC}, where the dark blue and yellow curves give the best overall description of the data, while the scenarios with $\lambda(x_p) \leq 0.7$
overestimate the deviation of the ratio from unity.

As can be seen from panel (a) of Fig.~\ref{fig:P_ZDC}, the data in the interval $E_{\rm ZDC} <  50$ TeV are best described, when $0.7 < \lambda(x_p) < 1$, while data in the $E_{\rm ZDC} >  150$ TeV range require that $\lambda(x_p) < 0.7$. This illustrates one of limitations of our model.
In particular, the region of very large $E_{\rm ZDC}$ corresponds to the large-$\nu$ tail of the distribution $P(\nu)$, which 
appears to be overestimated in our model.
A possible explanation of it could be that in the limit of a very large number of wounded nucleons $\nu$, one needs to take into account
the effect of energy-momentum conservation in cutting multiple Regge exchanges, which should suppress $P(\nu)$. 
This is neglected in our model, which assumes independent wounded nucleons.

A separate issue is re-interactions of hadrons produced in collisions with individual nucleons, which require additional modeling.
In particular, for central collisions and high multiplicities, while passing through a nucleus, the projectile proton
starts forming hadrons with soft momenta (in the nucleus rest frame) in a narrow tube-like volume in the longitudinal 
direction.
Since the acceptance of a ZDC in the transverse plane above 200 MeV/c is low, even very soft interactions of neutrons with
surrounding hadrons, which may give them a kick in the transverse direction, are likely to push those neutrons out of the ZDC acceptance.
This should lead to the suppression of the $E_{\rm ZDC}$ spectrum for large $E_{\rm ZDC}$, which is not included in our model.

It should be possible to refine our model once the data on neutron multiplicity for $\langle \nu \rangle =1-2$ in $\gamma A$ scattering becomes available. In particular, the $\langle \nu \rangle =1$ distribution can be extracted from dijet 
photoproduction at $x \geq 0.1$.

\section{Conclusions}
\label{sec:conclusions}

In this paper, we presented a model for the distribution of the number of forward neutrons $P_{\rm comb}(N)$ emitted in soft (minimum bias) and hard inelastic proton-nucleus scattering in the LHC kinematics.
It is based on the Gribov-Glauber model for the distribution $P(\nu)$ of the number of inelastic collisions
(wounded nucleons) $\nu$ in $pA$ scattering, where the composite hadronic structure of the projectile proton is included through its cross section (color) fluctuations, which in general depend on the active parton momentum fraction $x_p$. We examined and discussed main features of 
$P(\nu)$ and pointed out that the peak of $P(\nu)$ shifts toward lower $\nu$ with increasing $x_p$.

Assuming that each $\nu$ independently corresponds to the emission of $\langle M_n \rangle=5$ forward neutrons, 
we built a model for $P_{\rm comb}(N)$ and discussed its shape and dependence on $x_p$.
As an application, we converted $P_{\rm comb}(N)$ into the neutron energy distribution and showed that our model
provides a qualitative description of the trend of the ATLAS data on the ZDC energy spectra of forward neutrons emitted
in dijet production in inelastic $pA$ scattering at $\sqrt{s_{NN}}=8.16$ TeV.

In the future, one ultimately hopes to build a model for a centrality trigger, which would use the
ZDC signal to effectively separate 
the regimes of low and high nuclear density and combine proton-nucleus and photon-nucleus collisions; see~\cite{Alvioli:2024cmd} for 
a discussion of possible strategies in the $\gamma A$ case.

\acknowledgments
The authors would like to thank R.~Longo for useful discussions and comments on the manuscript.
The research of V.G.~was funded by the Academy of Finland project 330448, the Center of Excellence in Quark Matter
of the Academy of Finland (projects 346325 and 346326), and the European Research Council project ERC-2018-ADG-835105 YoctoLHC.
 The research of M.S.~was supported by the US Department of Energy Office
of Science, Office of Nuclear Physics under Award No. DE- FG02-93ER40771.


\begin{thebibliography}{99}

\bibitem{Salgado:2011wc}
C.~A.~Salgado, J.~Alvarez-Muniz, F.~Arleo, N.~Armesto, M.~Botje, M.~Cacciari, J.~Campbell, C.~Carli, B.~Cole and D.~D'Enterria, \textit{et al.}
J. Phys. G \textbf{39}, 015010 (2012)
[arXiv:1105.3919 [hep-ph]].

\bibitem{Citron:2018lsq}
Z.~Citron, A.~Dainese, J.~F.~Grosse-Oetringhaus, J.~M.~Jowett, Y.~J.~Lee, U.~A.~Wiedemann, M.~Winn, A.~Andronic, F.~Bellini and E.~Bruna, \textit{et al.}
CERN Yellow Rep. Monogr. \textbf{7}, 1159-1410 (2019)
[arXiv:1812.06772 [hep-ph]].

\bibitem{Baltz:2007kq}
A.~J.~Baltz, G.~Baur, D.~d'Enterria, L.~Frankfurt, F.~Gelis, V.~Guzey, K.~Hencken, Y.~Kharlov, M.~Klasen and S.~R.~Klein, \textit{et al.}
Phys. Rept. \textbf{458}, 1-171 (2008)
[arXiv:0706.3356 [nucl-ex]].

\bibitem{Contreras:2015dqa}
J.~G.~Contreras and J.~D.~Tapia Takaki,
Int. J. Mod. Phys. A \textbf{30}, 1542012 (2015)

\bibitem{Klein:2019qfb}
S.~R.~Klein and H.~M\"antysaari,
Nature Rev. Phys. \textbf{1}, no.11, 662-674 (2019)
[arXiv:1910.10858 [hep-ex]].

\bibitem{AbdulKhalek:2021gbh}
R.~Abdul Khalek, A.~Accardi, J.~Adam, D.~Adamiak, W.~Akers, M.~Albaladejo, A.~Al-bataineh, M.~G.~Alexeev, F.~Ameli and P.~Antonioli, \textit{et al.}
Nucl. Phys. A \textbf{1026}, 122447 (2022)
[arXiv:2103.05419 [physics.ins-det]].

\bibitem{Klasen:2023uqj}
M.~Klasen and H.~Paukkunen,
Ann. Rev. Nucl. Part. Sci. \textbf{74}, 49-87 (2024)
[arXiv:2311.00450 [hep-ph]].

\bibitem{Gribov:1968jf}
V.~N.~Gribov,
Sov. Phys. JETP \textbf{29}, 483-487 (1969) [Zh. Eksp. Teor. Fiz. \textbf{56}, 892-901 (1969)]

\bibitem{Frankfurt:2000tya}
L.~Frankfurt, V.~Guzey and M.~Strikman,
J. Phys. G \textbf{27}, R23-146 (2001)
[arXiv:hep-ph/0010248 [hep-ph]].

\bibitem{Frankfurt:2022jns}
L.~Frankfurt, V.~Guzey, A.~Stasto and M.~Strikman,
Rept. Prog. Phys. \textbf{85}, no.12, 126301 (2022) 
[arXiv:2203.12289 [hep-ph]].

\bibitem{Alvioli:2014sba}
M.~Alvioli, L.~Frankfurt, V.~Guzey and M.~Strikman,
Phys. Rev. C \textbf{90}, 034914 (2014)
[arXiv:1402.2868 [hep-ph]].

\bibitem{Alvioli:2014eda}
M.~Alvioli, B.~A.~Cole, L.~Frankfurt, D.~V.~Perepelitsa and M.~Strikman,
Phys. Rev. C \textbf{93}, no.1, 011902 (2016)
[arXiv:1409.7381 [hep-ph]].

\bibitem{Alvioli:2017wou}
M.~Alvioli, L.~Frankfurt, D.~Perepelitsa and M.~Strikman,
Phys. Rev. D \textbf{98}, no.7, 071502 (2018)
[arXiv:1709.04993 [hep-ph]].

\bibitem{ATLAS:2014cpa}
G.~Aad \textit{et al.} [ATLAS],
Phys. Lett. B \textbf{748}, 392-413 (2015)
[arXiv:1412.4092 [hep-ex]].

\bibitem{PHENIX:2015fgy}
A.~Adare \textit{et al.} [PHENIX],
Phys. Rev. Lett. \textbf{116}, no.12, 122301 (2016)
[arXiv:1509.04657 [nucl-ex]].

\bibitem{Frankfurt:1994hf}
L.~L.~Frankfurt, G.~A.~Miller and M.~Strikman,
Ann. Rev. Nucl. Part. Sci. \textbf{44}, 501-560 (1994)
[arXiv:hep-ph/9407274 [hep-ph]].

\bibitem{Dutta:2012ii}
D.~Dutta, K.~Hafidi and M.~Strikman,
Prog. Part. Nucl. Phys. \textbf{69}, 1-27 (2013)
[arXiv:1211.2826 [nucl-th]].

\bibitem{Burkardt:2002hr}
M.~Burkardt,
Int. J. Mod. Phys. A \textbf{18}, 173-208 (2003)
[arXiv:hep-ph/0207047 [hep-ph]].


\bibitem{Miller:2022kxt}
G.~A.~Miller,
MDPI Physics \textbf{4}, no.2, 590-596 (2022)
[arXiv:2203.02019 [hep-ph]].

\bibitem{Brodsky:2022bum}
S.~J.~Brodsky and G.~F.~de Teramond,
MDPI Physics \textbf{4}, no.2, 633-646 (2022)
[arXiv:2202.13283 [hep-ph]].

\bibitem{Frankfurt:2022tyy}
L.~Frankfurt and M.~Strikman,
MDPI Physics \textbf{4}, no.3, 774-786 (2022)

\bibitem{Alvioli:2024cmd}
M.~Alvioli, V.~Guzey and M.~Strikman,
Phys. Rev. C \textbf{110}, no.2, 025205 (2024)
[arXiv:2402.19060 [hep-ph]].

\bibitem{Alvioli:2016gfo}
M.~Alvioli, L.~Frankfurt, V.~Guzey, M.~Strikman and M.~Zhalov,
Phys. Lett. B \textbf{767}, 450-457 (2017)
[arXiv:1605.06606 [hep-ph]].

\bibitem{Strikman:2005ze}
M.~Strikman, M.~Tverskoy and M.~Zhalov,
Phys. Lett. B \textbf{626}, 72-79 (2005)
[arXiv:hep-ph/0505023 [hep-ph]].

\bibitem{Zheng:2014cha}
L.~Zheng, E.~C.~Aschenauer and J.~H.~Lee,
Eur. Phys. J. A \textbf{50}, no.12, 189 (2014)
[arXiv:1407.8055 [hep-ex]].


\bibitem{ATLAS:2025hac}
G.~Aad \textit{et al.} [ATLAS],
[arXiv:2504.02638 [nucl-ex]].

\bibitem{Alver:2008aq}
B.~Alver, M.~Baker, C.~Loizides and P.~Steinberg,
[arXiv:0805.4411 [nucl-ex]].

\bibitem{Alvioli:2012pu}
M.~Alvioli, H.~Holopainen, K.~J.~Eskola and M.~Strikman,
PoS \textbf{QNP2012}, 172 (2012)
[arXiv:1206.5720 [hep-ph]].

\bibitem{Bozek:2019wyr}
P.~Bo\.zek, W.~Broniowski, M.~Rybczynski and G.~Stefanek,
Comput. Phys. Commun. \textbf{245}, 106850 (2019)
[arXiv:1901.04484 [nucl-th]].

\bibitem{Lonnblad:2021fyl}
L.~L\"onnblad,
Nucl. Phys. A \textbf{1005}, 121873 (2021)

\bibitem{Bertocchi:1976bq}
L.~Bertocchi and D.~Treleani,
J. Phys. G \textbf{3}, 147 (1977)

\bibitem{Abramovsky:1973fm}
V.~A.~Abramovsky, V.~N.~Gribov and O.~V.~Kancheli,
Yad. Fiz. \textbf{18}, 595-616 (1973) [Sov. J. Nucl. Phys. \textbf{18}, 308-317 (1974)]

\bibitem{Good:1960ba}
M.~L.~Good and W.~D.~Walker,
Phys. Rev. \textbf{120}, 1857-1860 (1960)

\bibitem{Miettinen:1978jb}
H.~I.~Miettinen and J.~Pumplin,
Phys. Rev. D \textbf{18}, 1696 (1978)

\bibitem{Blaettel:1993ah}
B.~Blaettel, G.~Baym, L.~L.~Frankfurt, H.~Heiselberg and M.~Strikman,
Phys. Rev. D \textbf{47}, 2761-2772 (1993)

\bibitem{Alvioli:2013vk}
M.~Alvioli and M.~Strikman,
Phys. Lett. B \textbf{722}, 347-354 (2013)
[arXiv:1301.0728 [hep-ph]].

\bibitem{Alvioli:2009ab}
M.~Alvioli, H.~J.~Drescher and M.~Strikman,
Phys. Lett. B \textbf{680}, 225-230 (2009)
[arXiv:0905.2670 [nucl-th]].

\bibitem{Hammelmann:2019vwd}
J.~Hammelmann \textit{et al.} [SMASH],
Phys. Rev. C \textbf{101}, no.6, 061901 (2020)
[arXiv:1908.10231 [nucl-th]].

\bibitem{Alvioli:2018jls}
M.~Alvioli and M.~Strikman,
Phys. Rev. C \textbf{100}, no.2, 024912 (2019)
[arXiv:1811.10078 [hep-ph]].


\bibitem{ParticleDataGroup:2018ovx}
M.~Tanabashi \textit{et al.} [Particle Data Group],
Phys. Rev. D \textbf{98}, no.3, 030001 (2018)

\bibitem{Alvioli:2010yk}
M.~Alvioli and M.~Strikman,
Phys. Rev. C \textbf{83}, 044905 (2011)
[arXiv:1008.2328 [nucl-th]].

\bibitem{Alvioli:2019kcy}
M.~Alvioli, M.~Azarkin, B.~Blok and M.~Strikman,
Eur. Phys. J. C \textbf{79}, no.6, 482 (2019)
[arXiv:1901.11266 [hep-ph]].

\bibitem{Guzey:2005tk}
V.~Guzey and M.~Strikman,
Phys. Lett. B \textbf{633}, 245-252 (2006)
[arXiv:hep-ph/0505088 [hep-ph]].

\bibitem{E665:1995utr}
M.~R.~Adams \textit{et al.} [E665],
Phys. Rev. Lett. \textbf{74}, 5198-5201 (1995)
[erratum: Phys. Rev. Lett. \textbf{80}, 2020-2021 (1998)]


\end{thebibliography}
\end{document}